\def\etal{{\rm et al. }}
\def\mpc{{\ \rm Mpc}}
\def\kpc{{\ \rm kpc}}
\def\kms{{\rm km\, s^{-1}}}

\documentclass[usenatbib]{mn2e}

\usepackage{amssymb}
\usepackage{graphicx}
\usepackage{bmpsize}
\usepackage{cite}
\usepackage{natbib}
\usepackage{hyperref} 
\usepackage{aas_macros}
\usepackage{times}
\usepackage{amsmath}
\usepackage{natbib}
\usepackage{longtable}
\usepackage{txfonts}
\usepackage{pdflscape}
\usepackage{subfigure}
\usepackage{xcolor} 

\begin{document}

\title{On the nature of Small Galaxy Systems}
\author[Duplancic \etal]{Fernanda Duplancic$^{1}$\thanks{E-mail: fduplancic@unsj-cuim.edu.ar}, Georgina V. Coldwell$^{1}$, Sol Alonso$^{1}$ \& Diego G. Lambas$^{2}$\\ 
$^{1}$ Departamento de Geof\'{i}sica y Astronom\'{i}a, CONICET, Facultad de Ciencias Exactas, F\'{i}sicas y Naturales, Universidad Nacional \\
de San Juan, Av. Ignacio de la Roza 590 (O), J5402DCS, Rivadavia, San Juan, Argentina\\
$^{2}$ Instituto de Astronom\'ia Te\'orica y Experimental (IATE-CONICET), Laprida 854, C\'ordoba, Argentina }

\date{\today}

\pagerange{\pageref{firstpage}--\pageref{lastpage}}

\maketitle

\label{firstpage}

\begin{abstract}

We aim at defining homogeneous selection criteria of small galaxy systems in order to build catalogues suitable to compare main properties of pairs, triplets, and groups with four or more members. 
To this end we use spectroscopic and photometric SDSS data to identify systems with a low number of members. We study global properties of these systems and the properties of their member galaxies finding that  galaxies in groups are systematically redder and with lower star formation activity indicators than galaxies in pairs which have a higher fraction of star forming galaxies. Triplet galaxies present intermediate trends between pairs and groups. We also find an enhancement of star formation activity for galaxies in small systems with companions closer than 100$\kpc$, irrespective the number of members. 
We have tested these analysis on SDSS mock catalogues derived from the Millennium simulation, finding as conservative thresholds $76\%$ completeness and a contamination of $23\%$ in small galaxy systems, when considering an extreme case of incompleteness due to fiber collisions. Nevertheless, we also found that the results obtained are not likely affected by projection effects.
Our studies suggest that an extra galaxy in a system modify the properties of the member galaxies.
In pairs, galaxy-galaxy interactions increases gas density and trigger starbursts. However, repeated interactions in triplets and groups can generate gas stripping, turbulence and shocks quenching the star formation in these systems.

\end{abstract}

\begin{keywords}
galaxies: groups: general;
galaxies: interactions;
galaxies: statistics
\end{keywords}

\section{Introduction}

In the standard scenario of hierarchical clustering in the Universe, the study of the environment is crucial to understand galaxy evolution. Several properties of galaxies as morphology, total stellar mass, colour and star formation indicators have a strong correlation with environment, so that galaxies populating denser environments usually present early-type morphologies and are brighter and redder than galaxies in low density regions \citep{Dressler1980,Kauffmann2004,OMill2008,Peng2010,Peng2014,Zheng2017}. There are also  direct links between mechanisms affecting  galaxies and the environment. For example ram pressure stripping where the  intergalactic gas is removed as a galaxy travels through the intracluster medium  \citep{GunnGott1972} or galaxy harassment induced by a rapid interaction with other galaxies \citep{Moore1996} are common in galaxy clusters. Tidal interactions and mergers between galaxies can also modify galaxy morphology \citep{Toomre1972}, and trigger starbursts depending on the gas reservoir of the galaxies \citep{Yee-Ellingson1995,Kennicutt1998}. In this context, compact groups of galaxies are excellent laboratories for the study of galaxy evolution, due to the proximity of their members and the low velocity dispersion of the system create a perfect scenario for galaxy-galaxy interactions and mergers. 

Current knowledge of small galaxy systems is mainly based on individual studies of properties of galaxy pairs, triple systems and groups with four or more members. In the literature, it can be found different catalogues with very different selection criteria. For instance in the identification of galaxy pairs, the selection is generally based on the identification of galaxies close in projected distance and radial velocity difference. Nevertheless, there are no particular values established for the maximum separation between galaxies in a pair, so that adopted values range from 25 $ \kpc $ up to 1 $\mpc $ for the projected distance, and for the radial velocity difference, a few hundred $ \kms $ to limits of 1000 $\kms$ \citep [eg,][]{Patton2000,Lambas2003,Lambas2012, Ellison2008,Scudder2012,Argudo2015}. However, the contamination by projected pairs always affects the sample selection, for example \citet{Soares2007} shows that more than half of the simulated pairs  with projected separations lower than $50\ \kpc $ have significantly larger three-dimensional distances.

Pioneering studies of triple galaxy systems were conducted by \citet{Karachentsev1988}. These authors consider a set of three galaxies, with major axis $ a_1 $, $ a_2 $ and $ a_3 $ in the range [0.5 $ a_1 $ -2 $ a_1$] as an isolated triplet if the distance to the next significant neighbours is greater than three times the average distance between the triplet galaxies, and if the relative radial velocity differences are lower than 500 $\kms $ for all its members. \citet{Trofimov1995} compiled a list of triple systems with an average separation between galaxies of 600 $\kpc$ and a radial velocity difference cut of 150 $\kms$. \citet{Elyiv2009} developed a geometric method based on a high order Voronoi mosaic to identify triple systems finding that triplets with harmonic radii smaller than 200 $\kpc$ are mostly isolated, and have a larger probability of being real physical systems. \citet{OMill2012} identified triple systems of bright galaxies with projected separation lower than 200 $\kpc$ and a difference in radial velocity lower than 700 $\kms$. These systems were isolated by requiring no neighbours in a fixed aperture of 500 $\kpc$ from the triplet centre. On the other hand \citet{Argudo2015} selected triple systems considering a separation of 450 $\kpc$ between member galaxies and a maximum difference in radial velocity of 160 $\kms$.

Several catalogues of loose galaxy groups constructed from different versions of the FOF code \citep {Huchra1982}, have been presented in the literature \citep [e.g.,][]{Merchan2005, Tago2010, Tempel2012, Tempel2016}. This method links galaxies close in both projected separation and radial velocity difference. All galaxies linked by a common neighbour form a group if the numerical overdensity contrast exceeds a given threshold. Regarding compact groups, most catalogues were based on Hickson criteria \citep{Hickson1982} and a variety of samples have been generated from different galaxy surveys \citep [e.g.,] [] {Hickson1992, Lee2004, McConnachie2009, Euge2012}. A compact group is formed by four or more close galaxies, with magnitude difference  lower than three and with no significant neighbours within an aperture of radius equal to three times the group radius. Besides, the system compactness is defined through the effective surface brightness of the group, and different constraints were imposed on this parameter to select compact systems. 

It is important to highlight that these studies are based on catalogues constructed under different selection criteria. For example, the isolation criteria for Hickson's compact group and the one used in \citet{OMill2012} to identify triplets, are quite restrictive in comparison to the most common specifications used to identify pairs of galaxies, where the systems may be immersed in larger structures such as clusters or groups of galaxies \citep[e.g.][]{Alonso2004,McIntosh2008,Ellison2010,Alonso2012}.

Motivated by the diversity of criteria in the identification procedure of galaxy systems with a low number of members, we  aim at establishing an homogeneous selection criterion for the identification of compact, small galaxy systems, relatively isolated and with at least two member galaxies. These conditions foster  galactic interactions that could drive strong dynamical evolution. This catalogue will be suitable for different statistical studies of small galaxy systems and their member galaxies undergoing different astrophysical processes that affect the properties of galaxies.

This paper is organised as follows: in section \ref{data} we describe the  data used in this work. In section \ref{catalogue} we present the selection criteria for the construction of the small galaxy systems sample and a study of completeness and contamination by using mock catalogues. A study of the global properties of these systems is detailed in section \ref{global} and in section \ref{prop} we explore the main properties of galaxies in small systems. Finally in section \ref{conc} we present our main results.

Throughout this paper we adopt a cosmological model characterised by the parameters $\Omega_m=0.3$, $\Omega_{\Lambda}=0.7$ and $\rm H_0=70~h~{\rm km~s^{-1}~Mpc^{-1}}$.

\section{Data}
\label{data}

The sample of galaxies were drawn from the Data Release 14 of Sloan Digital Sky Survey\footnote{https://www.sdss.org/dr14/} \citep[SDSS-DR14,][]{Abolfathi2018,Blanton2017}. This survey includes imaging in 5 broad band ($ugriz$), reduced and calibrated using the final set of SDSS pipelines. The SDSS-DR14 provides spectroscopy of roughly two millions extragalactic objects including objects from 
the SDSS-I/II Legacy Survey \citep{Eisenstein2001,Strauss2002}, the Baryon Oscillation Spectroscopic Survey \citep[BOSS,][]{Dawson2013} and the extended-BOSS  \citep[eBOSS][]{Dawson2016}, from SDSS-III/IV.

In this work we consider Legacy survey area and obtain all data catalogues through $\rm SQL$ queries in $\rm CasJobs$\footnote{http://skyserver.sdss.org/casjobs/}. We select extinction corrected model magnitudes which are more appropriated for extended objects and also provide more robust galaxy colours. The magnitudes are k-corrected using the empirical k-corrections presented by \citet{OMill2011}. We restrict our analysis to galaxies with $r$-band petrosian apparent magnitude in the range $13.5<r<17.77$. The lower limit is chosen in order to avoid saturated stars in the sample and the upper limit corresponds to the limiting magnitude of the spectroscopic SDSS Main Galaxy Sample \citep[MGS,][]{Strauss2002}.

We consider two samples:
\begin{itemize}
\item Spectroscopic Galaxies: Objects with spectroscopic measurements in the Legacy survey in the redshift range $0.05<\rm z_{\rm spec}<0.15$
\item Photometric Galaxies: Objects with $ 13.5 <r <17.77 $ which have a distance to a spectroscopic galaxy lower than 55 arcsec but do not present spectroscopic measurements.  For these objects we consider photometric redshift in the range $0.01<\rm z_{\rm phot}<0.2$. 
\end{itemize}

The sample of photometric galaxies are used to assess the incompleteness due to fibre collision. In the SDSS spectrograph, fibres cannot be placed closer than 55 arcsec, therefore objects with a lower projected separation cannot be observed simultaneously. Due to this effect the SDSS spectroscopic sample is not fully complete. There are regions where the plates overlap (about 30\% of the mosaic), within close objects can be observed spectroscopically but the problem became relevant in the study of compact galaxy systems since 55 arcsec corresponds to $100\kpc$ at a redshift $z\sim 0.1$. Therefore in this study we used photometric redshift obtained from KF estimates stored in the table \texttt{Photoz} \citep{Beck2016}. We consider the following restrictions on the photometric redshift quality parameters, $\rm nnCount=100$, $\rm zErr>-1000$ and $\rm photoErrorCass=1$. For more information about KF quality parameters see SDSS documentation \footnote{http://www.sdss.org/dr14/algorithms/photo-z/}. 

Under these constraints we built a final parent galaxy sample comprising 422976 galaxies from which 97\% have spectroscopic measurements. 

\section{Small Galaxy Systems selection criteria}
\label{catalogue}

Several works on galaxy pairs suggest that, as expected, a large fraction of galaxies in close proximity in projection are undergoing galaxy-galaxy interactions \citep[e.g][]{Lambas2003,Lambas2012,Alonso2004,Alonso2006}. In order for an interaction to end in fusion, it is necessary to have both, proximity in 3-dim space and a low relative velocity between the galaxies, since fly-by encounters, frequent in high speed environments, are more likely to occur. Statistically speaking this means that the radial velocity difference between galaxies in a merger state will follow a Gaussian distribution, contrary to fly-by encounters where a flat distribution is expected.

Different works on the identification of small systems agree that a projected distance lower than 200$\kpc$ represents an appropriate scale to identify compact groups \citep{McConnachie2009,Elyiv2009,Duplancic2015}. 
For the case of pairs of galaxies, \citet{Patton2013} found a clear increase in the star formation activity of member galaxies up to projected separations of $150\kpc$, showing that the interaction affect galaxies up to these distances.

Regarding the velocity difference of group members, in the identification of merging systems the values range from about 150 $\kms$ to 1000 $\kms$. For instance, in order to consider galaxies physically associated within compact groups \citet{McConnachie2009} states that the maximum line-of-sight velocity difference between group members must be lower than 1000 $\kms$. On the other hand in the identification of galaxy pairs \citet{Alonso2006} and \citet{Lambas2012} use a radial velocity difference $\rm \Delta V< 350\ \kms$ and for galaxy triplets \citet{OMill2012} use $\rm \Delta V<700\ \kms$ while \citet{Argudo2015} consider $\rm \Delta V < 160\ \kms$ to select bound isolated pairs and triplets.

In order to identify small galaxy systems, as environments that promote mergers between galaxies, we consider galaxies with projected separation $r_p \le 200\kpc$ which is a suitable value to unify the projected distance selection criterion in the identification of systems with two or more members.
For the difference in radial velocity, it is important to choose low values since we want to identify systems which promote galactic interactions and fusions.
Because we are using both spectroscopic as well as photometric redshifts, we study the impact of different radial velocity cuts on the completeness and contamination rates in our catalogue. A detailed description of these analysis by using mock catalogues will be presented on section \ref{CCtest}.

The radial velocity limit was set according to the following cases:

\begin{itemize}
\item If both galaxies have spectroscopy, the maximum radial velocity difference will be  $\Delta V_{\rm ss}= 500\ \rm \kms$. 
\item If a spectroscopic galaxy has a photometric companion closer than 55 arcsec with magnitude $r<17.77$ then probably that galaxy has not been observed spectroscopically due to the restriction of the fibre size of SDSS spectrograph. In these cases the maximum radial velocity difference will be computed as  $\Delta V_{\rm sp}=\sqrt{(\Delta V_{\rm ss})^2+(c\ 1.5\ \rm \sigma_z)^2)}$ were $\rm \sigma_z$ is the error associated with the photometric redshift of the galaxy without spectroscopy and $c$ is the speed of light.
\item If both galaxies have no spectra, then the limit is estimated as $\Delta V_{\rm pp}=\sqrt{(c\ 1.5\ \rm \sigma_{zi})^2+(c\ 1.5\ \rm \sigma_{zj})^2)}$ were $\rm \sigma_{zi}$ and $\rm \sigma_{zj}$ are the error of the i galaxy and its neighbour j, respectively.
\end{itemize}

Moreover, in order to construct a sample of isolated systems we define a local isolation criterion which considers that there cannot be significant neighbours within a fixed aperture of $500 \kpc$ projected radius centred in the geometric centre of the group and radial velocity difference $\Delta V_{\rm iso-\rm spec} \leq 700 \rm km\ s^{-1}$ if the neighbour have spectroscopic information and  $\Delta V_{\rm iso- \rm phot}<\sqrt{(\Delta V_{\rm iso- \rm spec})^2+(c\ 1.5\ \rm \sigma_z)^2)}$ if it have photometric redshift. The velocity difference is calculated with respect to the average redshift of the group estimated only with its spectroscopic members. As significant neighbours we consider galaxies with absolute magnitude in the $r$ band brighter than $\rm M_r=-19$.

Under these constraints we construct a sample of 15963 systems comprising 13757 pairs, 1874 triplets and 332 groups with four or more members (hereafter groups). 

It is worth to notice that allowing fainter galaxies in the sample can introduce systems with satellites orbiting a dominant galaxy, therefore the dynamics of the group may be driven by the evolution of this bright galaxy. Therefore the relative luminosities between group members  could be of importance for the dynamical interpretation of the system. In order to avoid satellite galaxies in our sample we impose the restriction that the difference between the r-band absolute magnitude of the brightest and the faintest galaxy in the system should be lower than 2 magnitudes. This limit is consistent with the threshold defined by different author to select satellite galaxies \citep{Sales2005,AgustssonBrainerd2010,Lares2011}. Also \citet{Duplancic2015} performed a dynamical study of triplets of bright galaxies fulfilling this magnitude difference cut and found that the members reside in a single halo, suggesting that the systems have probably undergone recent merger events. From the original sample, 96\% of the systems fulfil this restriction, and the percentage of pairs, triplets and groups equal to 97\%, 90\% and 78\% respectively.

Our aim is to select compact systems therefore we also calculate the compactness parameter $S= \sum_{\rm i=1}^{\rm N} r_{\rm 90}^2\ /\ {R^2}$  were $r_{\rm 90}$ is the radius enclosing $90\%$ of the Petrosian flux of the galaxy in the r-band,  $R$ is the radius of the smallest circle containing the positions of member galaxies and N is the total number of members in the system. This parameter defined by \citet{Duplancic2013} is a measure of the percentage of the system total area that is filled by the light of member galaxies. In order to identify compact systems we require $S>0.03$, which corresponds to the compactness of compact groups of galaxies \citep{Duplancic2015}. Adding this restriction to the magnitude difference cut we retain 70\% of the systems, 73\% of the pairs, 42\% of triplets and only 30\% of groups. 

As a last constraint and in order to give confidence to the identification of the system as a physical entity, we restrict the selection to systems that have at least half of its members with spectroscopic measurements, loosing only 1\% of the systems.

Considering all these restrictions, our final sample  comprises 10929 small galaxy systems, from which 10044 are pairs, 791 triples and 94 groups. It is worth to notice that 80\% of the systems have all galaxy members with spectroscopy. Also 90\% of galaxy groups have 4 members and the richest systems have only 6 members, therefore the selection criteria is suitable to identify galaxy systems with a low number of members.  

\begin{figure*}
  \centering
  \includegraphics[width=.9\textwidth]{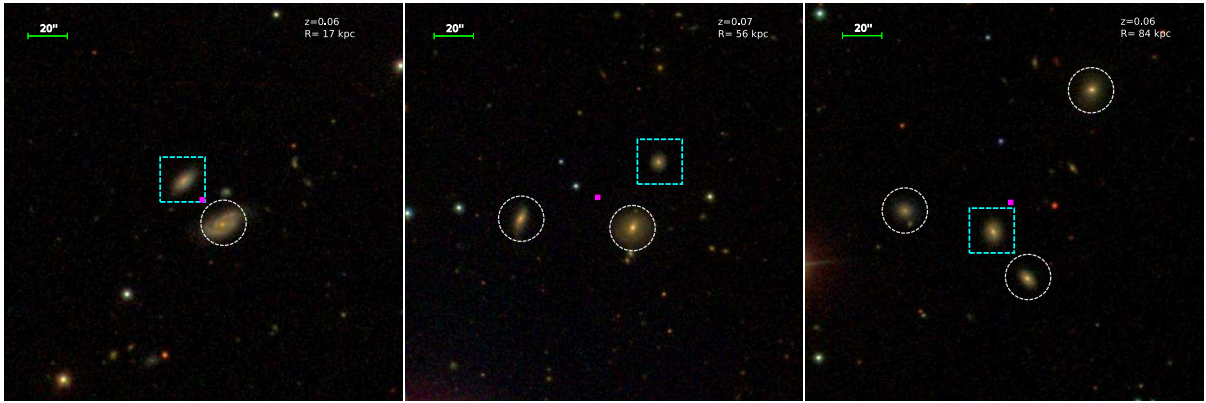}
\caption{Example of pair (left), triplet (middle) and groups (right) small galaxy systems. The dashed circles mark the position of spectroscopic member galaxies while photometric members are shown inside a dashed square. The squared dot represents the position of the geometric centre of the system. } 
\label{ej}
\end{figure*}

Fig. \ref{ej} shows example images of systems  with one photometric galaxy at z$\sim$0.06. It can be appreciated that the distances between member galaxies are of the order of their size and also that the systems present high degreed of isolation at least inside the image field which corresponds to about 250x250 $\kpc$ at the system redshift.

\subsection{Completeness and Contamination analysis}
\label{CCtest}
In order to test our selection criteria of small galaxy systems  we performed a completeness and contamination analysis by using mock SDSS light cones available at TAO\footnote{https://tao.asvo.org.au/tao/}. We select one random light cone based on Millennium simulation \citep{Springel2005} and SAGE semi-analytic model  \citep[2016 version,][]{Croton2016}. Millennium simulation uses WMAP-1 cosmology, has a box size of $500\ \rm h^{-1}\rm Mpc$, the mass resolution is $8.6\ 10^8 \rm h^{-1} \rm M_\odot$ and thee force resolution  $5\ \rm h^{-1} \rm kpc$. For the spectral energy distribution we select a Chabrier Initial mass function \citep{Conroy2009} and the Slab dust model of \citet{Devriendt1999}. The SDSS $ugriz$ magnitudes are provided in the AB system. To reproduce the survey geometry  we use \texttt{mangle}\footnote{http://space.mit.edu/~molly/mangle/} \citep{Swanson2008} with the window and mask \texttt{polygon} files provided by NYU-VAGC\footnote{ics.nyu.edu/vagc/} \citep{Blanton2005}.
The mock catalogue provides redshifts and angular positions with r-band apparent and absolute magnitude distributions similar to observational data, comprising 481018 mock galaxies with 40 different synthetic properties. 

In order to build a galaxy sample similar to the observational data used to identify small galaxy systems we consider the impact of fibre collision by including photometric galaxies into the mock catalogue. To this end we randomly select one galaxy in each pair closer than 55 arcsec and generate a photometric redshift by given an error to the observational redshift randomly selected from a Gaussian distribution that reproduces the photometric redshift error distribution obtained from the SDSS \texttt{Photoz} table. We set a redshift error to one galaxy in every close pair, which is a higher percentage than in the observational data and so represents an extreme case of incompleteness.

On these data we generate three different catalogues:
\begin{itemize}
\item R systems: Systems identified using cosmological redshift.
\item M systems: Systems identified using observed redshift.
\item Mz systems: Systems identified using observed plus photometric redshift.
\end{itemize}

In order to study redshift-space effects we use R and M systems identified using a projected distance cut $r_p= 200 \kpc$ and four different cuts in the radial velocity difference, $\Delta V_{\rm cut}=400\ \kms$, $500\ \kms$, $600\ \kms$ and $700\ \kms$. We also consider the same isolation criteria and magnitude difference cut used to identified small galaxy systems in the observational data. 

\begin{figure}
  \centering
  \includegraphics[width=.43\textwidth]{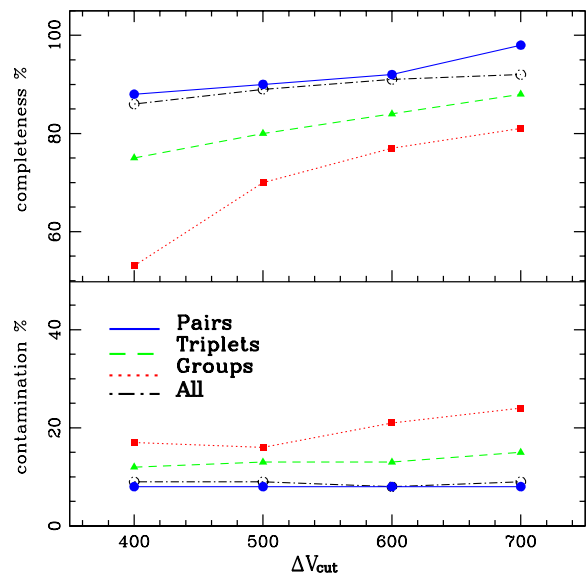}
\caption{Completeness and contamination rates as a function of radial velocity cuts, for pairs (solid), triplets (dashed), groups (dotted) and the entire sample of small galaxy systems (dot-dashed). } 
\label{comcont}
\end{figure}

We define completeness as the fraction of R systems that are also identified in the M catalogue,  and contamination as the fraction of M systems that are projections, i.e. not identified in the R catalogue. In Fig. \ref{comcont} we show the completeness and contamination fractions, as a function of the different radial velocity cuts, for pairs, triplets and groups, and also for all the sample of mock small galaxy systems. From this figure it can be seen that $\Delta V_{\rm cut}=500\ \kms$ represents a good compromise between high completeness and low contamination mostly for groups. 

With this radial velocity cut, by using cosmological redshift we identify 10633 R systems, from which 9705 are pairs, 850 triplets and 78 groups. The catalogue of small systems identified using observed redshifts (M systems) comprises 10397 objects, 9549 pairs, 783 triplets and 69 groups.  We are able to recover 89\% of R systems, 90\% of the pairs, 80\% of the triplets and 70\% of the groups respectively. In the M catalogue, 9\% are fake systems identified in projection. For galaxy pairs we find  contamination at 8\%,  triplets 13\%,  and groups 16\%. 

We also calculate the compactness $S$ for mock systems by using the disk scale radius parameter of mock galaxies which presents a similar distribution than the $r_{90}$ observational parameter. After the compactness cut we found no differences in completeness nor contamination rates.

Finally we use R and Mz systems to study the influence of photometric galaxies in the completeness and contamination of our catalogues. To this end we identify systems considering a projected separations $r_p\leq 200\ \kpc$ and radial velocity difference $\Delta V \leq 500\ \kms$. For the photometric mock galaxies we use the definitions of radial velocity cuts described in section \ref{catalogue}. We test different factors of the photometric redshift error, by considering 1 $\rm \sigma_z$, 1.5 $\rm \sigma_z$ and 2 $\rm \sigma_z$. For these three values we found no significant differences in the completeness nor contamination rates obtaining  completeness at 76\% (77\% for pairs 73\% for triplets and 70\% for groups) and contamination at 23\% (pairs 22\%, triplets 30\% and groups 39\%). Therefore we select the factor 1.5 $\rm \sigma_z$ for the selection of observational small galaxy systems.

It is important to highlight that we include a higher fraction of photometric galaxies in the mock catalogue than in the observational data, so 36\% of Mz systems  have at least one galaxy with photometric redshift while this fraction equals to 20\% in the observational sample of small galaxy systems. Therefore, the completeness and contamination fractions obtained using the Mz catalogue can be considered upper limits.

\section{Global Properties of small galaxy systems}
\label{global}

\begin{figure}
  \centering
  \includegraphics[width=.37\textwidth]{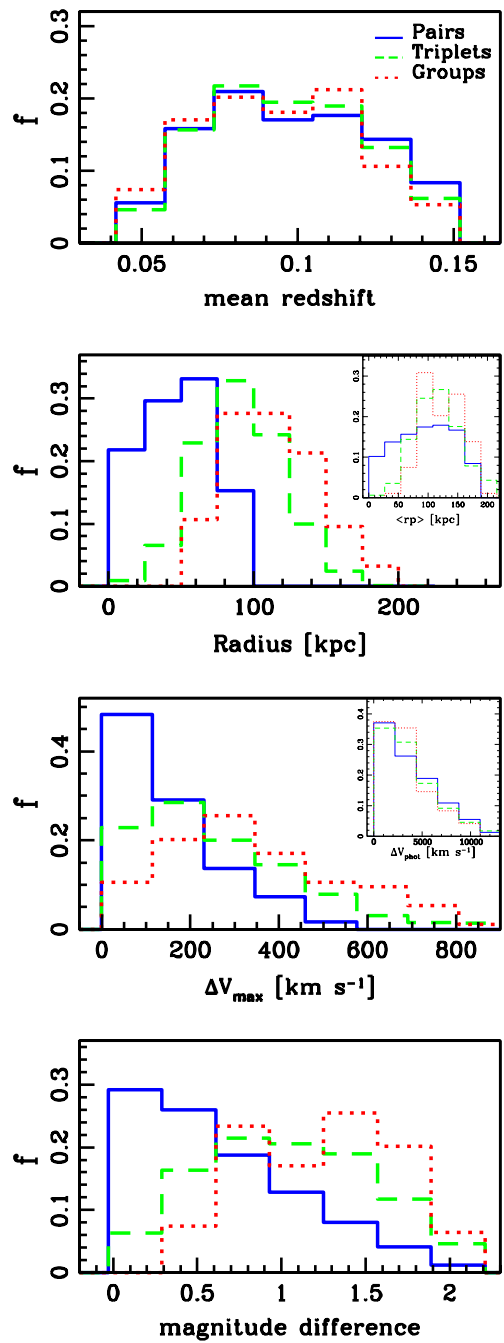}
\caption{Normalized distribution of the main properties small galaxy systems, distinguishing between pair (solid), triplet (dashed) and groups (dotted). From top to bottom, average redshift, radius of the system (the inset panel figure shows the average distance between galaxies), maximum radial velocity difference between spectroscopic galaxies in the system (the inset panel figure shows the velocity difference of photometric galaxies) and maximum magnitude difference between system members.} 
\label{groupsprop}
\end{figure}

In this section we explore the main characteristics of the sample of small galaxy systems constructed in this work. For pairs, triplets and groups, we estimate the system redshift as the average redshift of the galaxy members with spectroscopic measurements and the group radius as the projected distance from the group geometric centre to the most distant galaxy member. We also provide the maximum radial velocity difference between system spectroscopic members and the maximum absolute magnitude difference in the $r$ band between system members (Fig. \ref{groupsprop}). 

The redshift distribution (top panel Fig. \ref{groupsprop}) is similar for pairs, triplets and groups of our sample. A Kolmogorov-Smirnov (KS) test gives in all cases \texttt{p}$>$0.05 for the null hypothesis that the samples are drawn from the same distribution. Therefore, the results of our analysis regarding differences in the systems properties  are not likely to be biased by redshift effects. 

The system radius distribution is shown in the second panel of Fig. \ref{groupsprop} where it can be seen an increase from pairs, to triplets and groups. Nonetheless, the average separation between member galaxies span a similar range for all the systems although galaxies in pairs have a tendency to be closer, as can be seen in the inset figure of this panel. This result indicates that, on average, galaxies in small systems have similar relative separations within each other.

We computed the maximum radial velocity difference using only the spectroscopic members of the systems. In the third panel of Fig. \ref{groupsprop} it can be seen that the sample of galaxy pairs presents the lowest velocity difference values as well as an increasing trend for triplets and groups. The inset of this panel shows the radial velocity difference between the group centre and the photometric galaxies. Although the values are large due to the uncertainties in photometric redshifts, for pairs, triplets and groups it is seen a decreasing trend with velocity difference. These distributions are consistent with galaxies in a merger state, since no trend is expected for randomly selected galaxies. 

We also estimate the maximum magnitude difference between galaxy members. In the bottom panel of Fig. \ref{groupsprop} can be appreciated that this value is higher for groups, than for triplets and pairs. Moreover the magnitude difference between pair galaxies suggests that most of pair systems in our sample will undergo major merger events.

Table \ref{table1} present the mean value for these quantities and their corresponding errors calculated from bootstrap resampling techniques \citep{Barrow1984}, these values suitably represent the trends observed in the distributions of the main properties of pairs, triplets and groups.

\begin{table}
\centering
\caption{Main Properties of the small galaxy systems. Sample name, number of systems (N), number of galaxy members (Ngx), Group Radius ($\rm R$), maximum radial velocity difference ($\Delta \rm V_{\rm max}$),  and maximum magnitude difference ($|\rm M_{\rm rN}-M_{\rm r1}|$).}
\begin{tabular}{lcccccc}
\hline\hline\noalign{\smallskip}
Name & N & Ngx &$\rm R$ & $\Delta \rm V_{\rm max}$ &  $|\rm M_{\rm rN}-M_{\rm r1}|$\\
 &&&$\rm kpc$ & $\rm kms^{-1}$ & mag \\
\hline\noalign{\smallskip}
Pairs & 10044 &20088  &47.63$\pm$0.24 & 103$\pm$1 &   0.64$\pm$0.01\\
Triplets & 791 & 2373  & 90.97$\pm$0.99 & 261$\pm$6&  1.03$\pm$0.02 \\
Groups & 94 & 381 &  112.37$\pm$3.39 & 349$\pm$20 & 1.24$\pm$0.05\\
\hline
\end{tabular}
\label{table1}
\end{table}

\begin{figure}
  \centering
  \includegraphics[width=.35\textwidth]{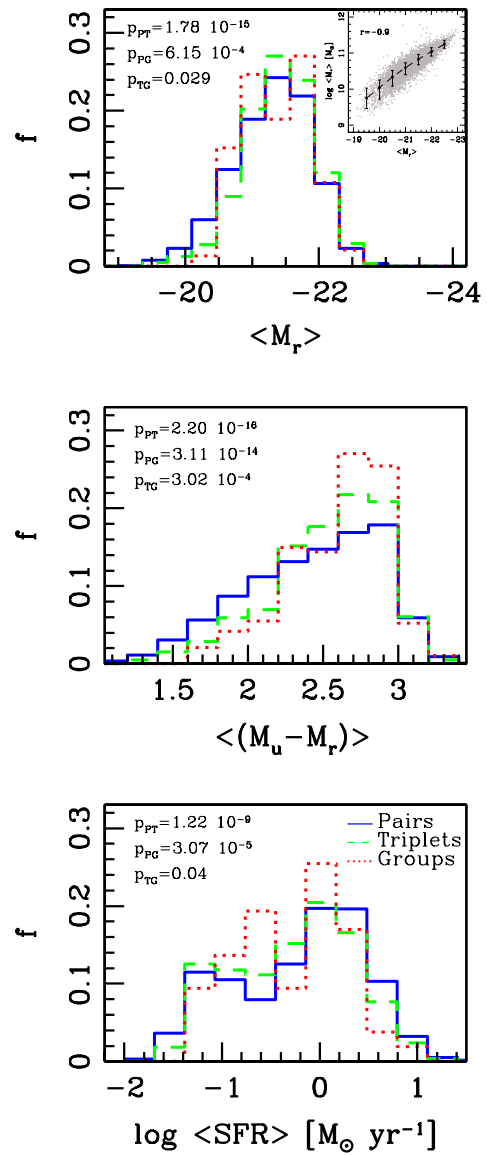}
\caption{Normalized distribution of the average system luminosity (top), the average colour index (middle) and the average star formation rate (bottom), for galaxy pairs (solid), triplets (dashed) and groups (dotted). The inset figure of top panel shows the relation between average luminosity and average stellar mass content for small galaxy systems with all their members with spectroscopic measurements along with the correlation coefficient. In this figure we also list $\texttt{p}$ values obtained from a KS test over pairs and triplets ($\rm p_{PT}$), pairs and groups ($\rm p_{PG}$) and triplets and groups ($\rm p_{TG}$).} 
\label{globales}
\end{figure}

As a complementary study, we explore the average luminosity of the systems calculated through the r-band absolute magnitudes of all members  in pairs, triplets and groups (top panel of Fig. \ref{globales}). We find that the average r-band luminosity is a suitable proxy of the average stellar mass content as these parameters are highly correlated (r=0.9). This can be appreciated in the inset figure of the top panel of Fig. \ref{globales} where we plot the average magnitude versus the average stellar mass content for systems with all their members with spectroscopic measurements. To compare the distribution of average luminosity of pairs, triplets and groups we performed a KS test finding in all cases $\texttt{p}<0.05$ therefore the distributions are expected to be different. 
We also study the average ($\rm M_u-\rm M_r$) colour and the average star formation rate of pairs, triplets and groups. In the middle and bottom panels of Fig. \ref{globales} it can be seen that groups have redder average colours ($\langle(\rm M_u-\rm M_r)\rangle >$2.4)  and a higher fraction of low star-forming systems (log $\langle \rm SFR \rangle<$-0.5) compared to triplets and pairs. These results are in agreement with the KS test giving in all cases $\texttt{p}<0.05$, therefore we reject the null hypothesis that the samples are drawn from the same distribution.

\subsection{The effect of fibre collision on system properties}
\label{fib}

In the present work we consider fibre collision effects following the methodology used in \citet{OMill2012} were the incompleteness due to fibre collision in the detection of triple systems  was studied by using photometric redshifts. The authors found that 90\% of the isolated spectroscopic triplets can be recovered using spectroscopic and photometric data, lost systems include triplets not fulfilling the isolation criteria due to the inclusion of bright galaxies without spectroscopy. New triplets identified with 1 or 2 members without spectroscopic information present a high degree of isolation and clear sign of interaction in their morphology such as bridges and tidal tails. 

\begin{figure}
  \centering
  \includegraphics[width=.35\textwidth]{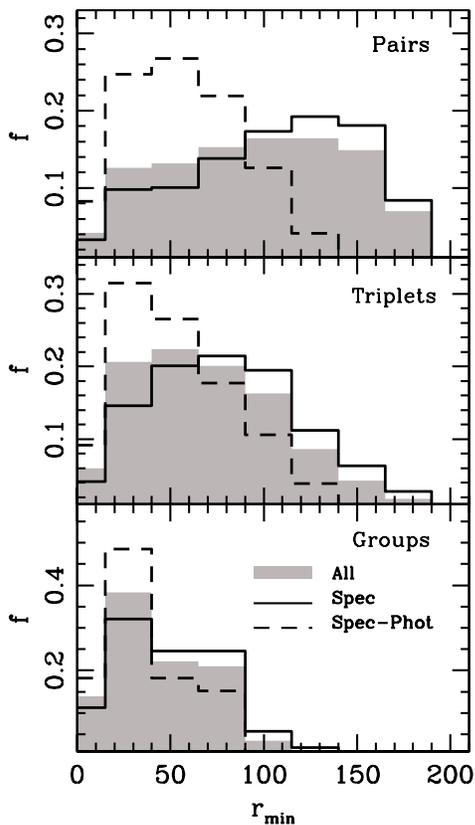}
\caption{Normalized distribution of the the minimum projected distance between group member galaxies ($\rm r_{\rm min}$) in the sample of pairs (top), triplets (middle) and groups (bottom). The shaded distribution corresponds to the full sample, in solid line we show the distribution of the systems that have all their members with spectroscopy (Spec) and in dashed line systems with at least one galaxy with photometric redshift (Spec-Phot).  } 
\label{fibcoll}
\end{figure}

In this section we explore the impact of including galaxies without spectroscopy due to fibre collision on the properties of small systems. Fig. \ref{fibcoll} shows the distribution of the the minimum projected distance between group member galaxies ($\rm r_{\rm min}$) in the sample  of small galaxy systems. We distinguish between pairs, triplets and groups. Also we consider separately the distribution of the systems that have all their member galaxies with spectroscopy (Spec) and those that have at least one galaxy with photometric redshift (Spec-Phot). 

We can see in this figure that the Spec-Phot systems have a distribution with lower values of $\rm r_{\rm min}$ compared to systems where all their members have spectroscopic determinations. Moreover the distribution of Spec-Phot systems drops for distances greater than $100 \kpc$ which is the physical size of the fibre at z=0.1. This is the expected trend because at distances greater than $100 \kpc$ two galaxies may be resolved spectroscopically. This tendency is stronger for galaxy pairs highlighting that the photometric data described in section \ref{data} is useful to complete the sample of close pairs.
These results are in agreement with the work of \citet{Mesa2014} who identify close tidal pairs by looking for photometric companions of spectroscopic galaxies. After a visual classification, these authors found that more than 60\% of spectro-photometric pairs have distorted morphologies with features such as tidal tails and bridges or clear signs of fusion in process.

\begin{table}
\centering
\caption{ Sample name, number(fraction) of systems (N),  systems Spec ($\rm N_{\rm Spec}$) and systems Spec-Phot ($\rm N_{\rm Spec-Phot}$).}
\begin{tabular}{lcccccc}
\hline\hline\noalign{\smallskip}
Name & N & $\rm N_{\rm Spec}$ &  $\rm N_{\rm Spec-Phot}$ \\
\hline\noalign{\smallskip}
Pairs & 10044 &8182 (0.82)  & 1862 (0.18)\\
Triplets & 791 & 508 (0.65)  & 283 (0.35)\\
Groups & 94 & 56 (0.60)  & 38 (0.40) \\
\hline
\end{tabular}
\label{table0}
\end{table}

Table \ref{table0} summarizes the total number of pairs, triplets and groups and the number of Spec and Spec-Phot systems in these samples. Galaxy pairs present the highest percentage of systems with all members galaxies with spectroscopy, reaching 80\%, while triplets and groups present a similar fraction of Spec systems of 65\% and 60\% respectively. It is worth to notice that as a consequence of the selection criteria, Spec-Phot triplets and pairs admit only one photometric member. For galaxy groups only 20\% of Spec-Phot systems have more than one photometric galaxy, being the maximum number of photometric members equal to two. Also the total fraction of spectroscopic galaxies in the sample is 90\%, therefore we conclude that the sample of small systems is dominated by galaxies with spectroscopic measurements.

\section{Analysis of the main properties of galaxies in small systems}
\label{prop}

\begin{figure}
  \centering
  \includegraphics[width=.37\textwidth]{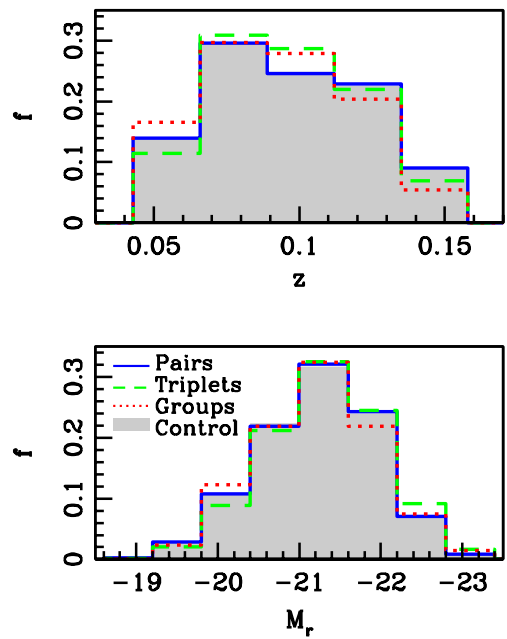}
\caption{Normalized distribution of redshift (top) and absolute magnitude in the r band (bottom) for spectroscopic galaxies in the sample of galaxy pairs (solid), triplets (dashed), groups (dotted) and the control sample (shaded).} 
\label{gxmainprop}
\end{figure}

Regarding to the properties of galaxies in small systems, it is known that groups present a higher fraction of red, early-type galaxies compared to the field population \citep[e.g.,][]{Hickson1988, Palumbo1995, Brasseur2009, Coenda2015}. On the other hand, galaxies in pairs exhibit a higher star formation rate and younger stellar populations compared to non-interacting galaxies \citep [e.g.,][]{Lambas2003, Alonso2004, Patton2011, Ellison2013, Behroozi2015}. For triple galaxy systems \citet{Duplancic2013} found that triplets of bright galaxies have properties similar to the members of compact groups. However, most of the galaxy catalogues used in these studies were built under different selection criteria. For instance, there are differences in the distance between member galaxies, radial velocity cuts and isolation constraints.

In this section we study the main properties of galaxies in pairs, triplets and groups identified under an homogeneous selection criteria of small galaxy systems. We consider only spectroscopic galaxies, but as the fraction of galaxies with spectra in our sample is 90\%, the results obtained in this section are likely to represent faithfully the trends in the total sample.

\begin{figure*}
  \centering
  \includegraphics[width=.9\textwidth]{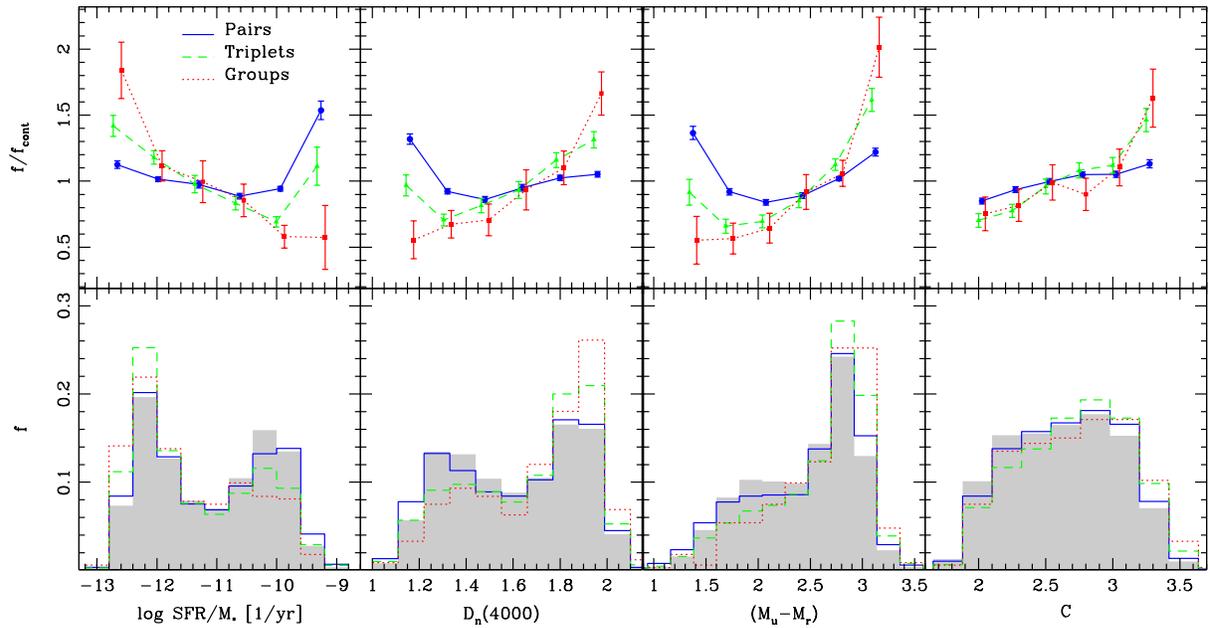}
\caption{From left to right, specific star formation rate $\rm SFR/M_*$, $\rm D_n(4000)$ index, ($\rm M_u-\rm M_r$) colour and concentration index $C$. Bottom: normalized distributions of spectroscopic galaxies in pairs (solid), triplets (dashed) and groups (dotted). The shaded distribution corresponds to galaxies in the control sample. Top: Quotient between the distributions of pairs (solid), triplets (dashed) and groups (dotted) and the control sample. The error bars correspond to standard uncertainties.} 
\label{gxprop}
\end{figure*}

For a suitable comparison of the properties of galaxies in small systems, we built a control sample by randomly selecting spectroscopic galaxies to match the redshift and  r-band absolute magnitude distributions of spectroscopic members in small systems. For these galaxies we also request no companions ($\rm M_r\leq -19$) within a projected separation of $200\kpc$ and a radial velocity difference of $500\ \kms$. This sample comprises 10 times more galaxies than the sample of galaxies in small systems.

Fig. \ref{gxmainprop} shows the distributions of redshift and r-band absolute magnitude of spectroscopic galaxies in the sample of pairs, triplets, groups and the control sample. As can be seen, the samples show  similar distributions ($\texttt{p}>0.05$, for a KS test), so that our results are not expected to be biased for differences in neither redshift nor magnitude of their galaxies.

In order to study the properties of galaxies in small systems we use the \texttt{galSpec} galaxy properties from MPA-JHU emission line analysis for the SDSS-DR8 \footnote{Available at http://www.sdss.org/dr12/spectro/galaxy\_mpajhu/}. From this catalogue we consider as a spectral indicator of the stellar population mean age the strength of the $4000$ \AA{} break ($\rm D_n(4000)$) defined as the ratio of the average flux densities in the narrow continuum bands 3850-3950 \AA{} and 4000-4100 \AA{} \citep{Balogh1999}. We also use the star formation rate (SFR) and specific star formation rates ($\rm SFR/M_*$) according to \citet{Brinchmann2004} and total stellar masses ($\rm M_*$) calculated from the photometry \citep{Kauffmann2003}.  It is important to highlight that although the data used in this study correspond to SDSS-DR14, the 99\% of spectroscopic galaxies in the sample of small systems have spectroscopic measurements in MPA-JHU data. We also consider the concentration index $\rm C=\rm r_{\rm 90}/\rm r_{\rm 50}$, where $\rm r_{\rm 90}$ and $\rm r_{\rm 50}$ are the radii containing 90\% and 50\% of the Petrosian galaxy light in the r band. 
This parameter is a suitable indicator of galaxy morphology: early type galaxies have $\rm C>2.6$ while late type galaxies have typically $\rm C<2.6$ \citep{Strateva2001}

In the bottom panels of Fig. \ref{gxprop} we explore the distribution of the specific star formation rate, $\rm D_n(4000)$ index as stellar population age indicator, ($\rm M_u-\rm M_r$) colour and concentration index C, for galaxies in pairs, triplets and groups.  The shaded areas in this figure represent the distributions of these parameters for galaxies in the control sample. A significant difference between the control sample and groups distributions is observed for the specific star formation rate $\rm SFR/M_*$, $\rm D_n(4000)$ index and colours of the galaxies. These discrepancies with the control sample become smaller for galaxies in pairs and triplets. Nonetheless, the morphological distribution of all the samples is rather similar.

In order to compare and quantify the differences of the properties of galaxies in small systems with respect to the control sample we consider 6 bins and divide the fraction of galaxies in pair, triplets and groups with the fraction of galaxies in the control sample (top panels Fig. \ref{gxprop}). 
It can be seen from this figure than only pairs present a higher fraction of star forming, young stellar population, and blue colour galaxies with respect to the control sample. Galaxies in groups present an opposite trend showing an excess of passive, red, old stellar population galaxies in comparison to the control sample. For triplets, the trends are intermediate but the fraction of passive galaxies is lower than for groups. Regarding morphology, galaxies in pairs are similar to galaxies in the control sample while triplet and group have a weak excess, within the errorbars, of galaxies with bulge type morphology, compared with the control sample.

\begin{figure}
  \centering
  \includegraphics[width=.3\textwidth]{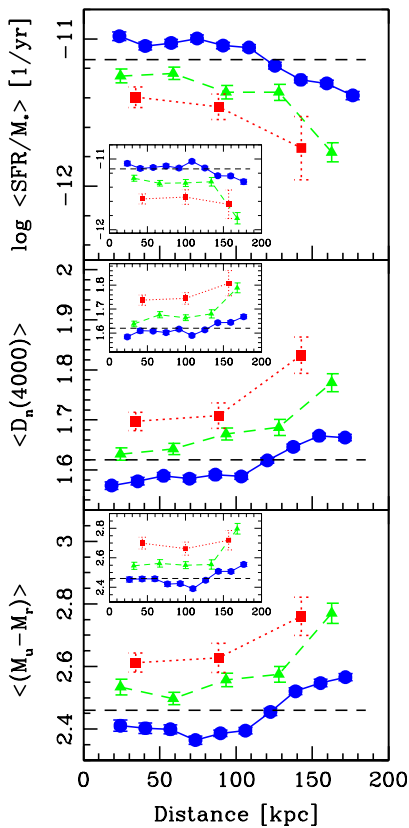}
\caption{Average values of specific star formation rate (top), $\rm D_n(4000)$ index (middle) and $(\rm M_u-\rm M_r)$ colour as a function of the distance to the nearest companion within the system, for galaxies in pairs (solid), triplets (dashed) and groups (dotted). Error bars were calculated using bootstrap resampling techniques. The horizontal dashed line represents the mean values of the parameters for spectroscopic galaxies in the control sample. The inset panels show the trend for a sample with no variation of redshift nor magnitude when dividing the sample in distance to the nearest companion closer/farther than $100 \kpc$ (see text).  } 
\label{proprpmin}
\end{figure}

In different works it was shown that the proximity between galaxies may drive burst of star formation \citep[e.g.][]{Lambas2003,Patton2013}. It is important to consider that the distance to the nearest member galaxy in the system may be affecting the main properties of galaxies. Therefore in Fig. \ref{proprpmin} we study the average specific star formation rate, stellar population ages represented in the $\rm D_n(4000)$ parameter and mean colour of galaxies in pairs, triplets and groups, as a function  of the distance to the nearest companion within the system. The horizontal dashed line corresponds to the mean values of the parameters for spectroscopic galaxies in the control sample. 

From this figure it can be observed that, for galaxies in small systems, there is a trend to increase the population of star forming, young stellar population, blue galaxies with decreasing distance to the nearest companion. Nevertheless, galaxies in pairs always present a higher star formation activity than galaxies in triplets and groups with significant differences (at more than 3 $\sigma$ level) between the properties of pairs and group galaxies, despite the distance to the nearest companion. For triplet members we find intermediate trends  namely, bluer, star forming and with younger stellar populations than galaxies in groups  (which are the redder objects and with the lowest star formation activity in the sample of small galaxy systems). Nevertheless, the differences between the properties of galaxies in triplets and groups are at less than $\sigma$ level in most cases.

An important result from this study is that for galaxies with companions closer than 100$\kpc$ there is a clear enhancement in the mean value of the specific star formation rate, $\rm D_n(4000)$ index and $(\rm M_u-\rm M_r)$ colour for all galaxies in small systems despite the number of members, but only galaxies in pairs achieve values corresponding to bluer, star forming and with a younger stellar population objects than the average values of the control sample. 

In order to check for systematics in these trends we consider galaxies with a companion closer/farther than $100\kpc$ and explore the redshift and absolute r-band magnitude distributions of these samples. We found slight differences in the distribution of z and $\rm M_r$, observing than for galaxies with companions at distances greater than $100\kpc$ the distributions shifts towards larger redshifts and brighter magnitudes. For this reason, we randomly select galaxies to match a reference sample with a similar distribution of these fundamental parameters. The resulting sample comprises the 83\% of the galaxies in pairs, triplets and groups. The inset panels of figure \ref{proprpmin} show the variation of the average specific star formation rate, stellar population ages and mean colour for pairs, triplets and groups in this sample, finding no significant differences with respect to the previously obtained trends.

\subsection{Influence of projections on galaxy properties}
\label{projgxs}

\begin{figure}
  \centering
  \includegraphics[width=.4\textwidth]{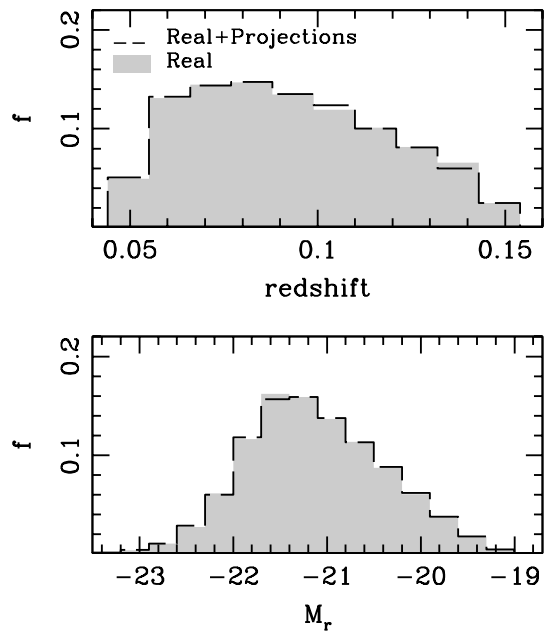}
\caption{Normalized distributions of redshift (top) and r-band absolute magnitude $\rm M_r$ (bottom) for galaxies in the \textit{ Real$+$projections} and \textit{ Real} sample. } 
\label{zMrmock}
\end{figure}

To study the influence of projections on the properties of spectroscopic galaxies populating small galaxy systems we use the Mz catalogue defined in section \ref{comcont} that comprises systems identified using photometric mock galaxies. On this data we consider two samples, one including all 16572 spectroscopic galaxy members of Mz systems (\textit{Real$+$projections}), and a second sample comprising 15447 galaxies populating real systems (identified using cosmological redshift) in the Mz catalogue (\textit{Real}). In Fig. \ref{zMrmock} we show the redshift and r-band absolute magnitude distributions of these samples. From this figure it can be appreciated that both quantities present equal trends ($\texttt{p}\sim 0.9$ for a KS test), therefore no systematic effects associated to differences in these distributions are expected.  

\begin{figure}
  \centering
  \includegraphics[width=.4\textwidth]{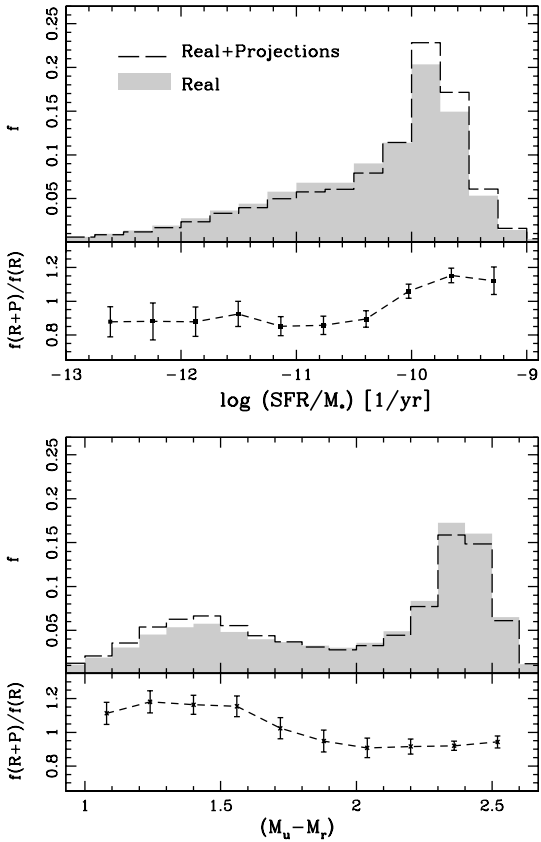}
\caption{Normalized distributions of specific star formation rate (top) and $(\rm M_u-\rm M_r)$ colour (bottom) along with the quotient between the fractions of galaxies in the \textit{ Real$+$projections} sample and the fraction of galaxies in the \textit{ Real} sample. The error-bars correspond to standard uncertainties. } 
\label{diffmock}
\end{figure}

In Fig. \ref{diffmock} we show the distributions of the specific star formation rate and $(\rm M_u-\rm M_r)$ colour of theses samples along with the ratio between the fraction of galaxies in the \textit{ Real$+$projections} sample and the fraction of galaxies in the \textit{ Real} sample, considering 10 bins of galaxy properties. 
From this figure it can be seen that galaxies in projection show an increase of the fraction of galaxies with higher specific star formation rate and bluer colours resulting in slightly differences in the mean values of these parameters: log $\langle \rm SFR/\rm M_*\rangle$=-10.36$\pm$0.01  and  $\langle (\rm M_u-\rm M_r)\rangle$=1.95$\pm$0.01 for  \textit{Real$+$projections} sample and log $\langle \rm SFR/\rm M_* \rangle$=-10.45$\pm$0.01 and and $\langle (\rm M_u-\rm M_r) \rangle$=1.99$\pm$0.01 for galaxies in the \textit{Real} sample. Nevertheless, the average excess is of the order of 10\%, so we conclude that projection effects do not significantly affect the results obtained from the analysis of galaxy properties in the sample of observational small systems.

\section{Conclusions}
\label{conc}

In this work we study main properties of galaxies in small systems. To this end we use spectroscopic and photometric data from SDSS-DR14 and define a homogeneous selection criteria to identify systems with a low number of members populating  environments that promote galaxy-galaxy interactions and mergers. We consider galaxies close in projection in the plane on the the sky and with low radial velocity differences. We locally isolate the system considering a fixed aperture restriction that prevent significant neighbours within 500~$\rm kpc$ and with a radial velocity cut of 700 $\kms$. We also consider system with a compacticity similar to Hickson compact groups, and populated by galaxies with similar luminosities. In order to give confidence to the identification of the system as a physical entity, we restrict the selection to systems that have at least half of their members with spectroscopic measurements. Under these constraints we construct a sample of 10929 small galaxy systems, from which 10044 are pairs, 791 triples and 94 groups with four or more members.

To study projection effects in our sample we use a SDSS mock catalogue and found conservative rates of 76\% completeness and 23\% contamination, when including an extreme case of incompleteness due to fibre collisions on SDSS data. The differences between the real and real+projected properties of galaxies are of the order of 10\% being the mean values of specific star formation rate and colour of galaxies in real systems unaltered by these projections.

By using photometric data we consider the effect of fibre collision in our sample of small systems finding that 80\% of galaxy pairs have all members galaxies with spectroscopy, while triplets present a fraction of systems with all spectroscopic members of 65\%. The remaining fraction of triplets and pairs have only one photometric member. Galaxy groups present a fraction of 60\% systems with all spectroscopic members  with only 20\% of spectrophotometric groups with more than one photometric galaxy and a maximum number of two photometric members. Also we find that 90\% of galaxies in small systems have spectroscopic measurements.

We study the global properties of these systems and found that the average separation between member galaxies span a similar range for pairs, triplets and groups, although galaxies in pair tend to be closer. Also we study the maximum radial velocity and magnitude difference between system members finding  an increasing trend from pairs to groups. The proximity, low relative velocity and magnitude difference between pair members suggest that these systems will undergo major merger events. 

We have also performed an analysis of the main properties of spectroscopic galaxies in small systems finding that galaxies in groups are systematically redder and with a lower star formation activity than galaxies in pairs which exhibit a higher fraction of star forming galaxies. For galaxies in triplets the tendency is similar than for groups although in these systems there is a fraction of galaxies with star formation rate and stellar population ages similar to galaxies in pairs. These results are in agreement with the work developed in \citet{Duplancic2013} who studied a sample of triplets of bright galaxies obtaining a small fraction of globally blue triplets.

We also considered the average specific star formation rate, the stellar population ages represented by the $\rm D_n(4000)$ parameter, and the mean colour of galaxies in pairs, triplets and groups, as a function  of the distance to the nearest companion within the system. We find that galaxies in pairs are more strongly star forming and bluer than galaxies in triplets and groups for all the range of separations. Moreover, only galaxies in pairs present values of star formation indicators higher than the average of SDSS spectroscopic galaxies.

We also find an enhancement of the mean value of the specific star formation rate, $\rm D_n(4000)$ index and $(\rm M_u-\rm M_r)$ colour for galaxies with companions closer than 100$\kpc$ in small systems despite the number of members. This may be an indication of an interaction-induced star formation activity due to recent close encounters.  

It should be noted that the analysis of the properties of galaxies in pairs, triplets and groups has been carried out in density environments locally controlled through an isolation criterion. Moreover, our sample of pairs, triplets and groups were built under a homogeneous selection criteria and have similar distributions of redshifts and absolute r-band magnitude. 

Our results suggest important effects of environment on the properties of galaxies, even at very small scales, where the difference of an extra galaxy in a system can significantly modify the properties of the rest of the member galaxies. Our sample of compact systems is isolated in a fixed aperture of 500$\kpc$ ($\rm \Delta V\leq 700 \ \kms$). When a galaxy residing in this isolated local environment has only one close companion ($\rm r_p\leq 200 \kpc$, $\rm \Delta V\leq 500 \ \kms$), the gravitational interaction between these two galaxies may generate an increase of the star formation activity tidally induced by close interactions. This effect was studied in the pioneering work of \citet{BarnesHernquist1996} with numerical simulations showing that during an interaction the gas component in galaxies can significantly lose angular momentum due to gravitational torques, increasing gas density and triggering a starburst. Also, in high-resolution  numerical simulations aimed to study galaxy-galaxy interactions  \citet{Teyssier2010} show that the dominant process triggering star burst is gas fragmentation into massive and dense clouds. In this line, \citet{Violino2018} investigate the effects of interactions on the molecular gas component of galaxies undergoing a merger finding that the  average  molecular  gas  consumption  timescale  of  galaxy pairs is shorter than in non-interacting galaxies. Therefore, as a consequence of the gravitational interaction with a close companion a faster transition of the molecular gas to its denser phase takes place, increasing the star formation efficiency.

Our results suggest that in triplets the efficiency of star formation decreases. Moreover, in systems with more than three members, star formation efficiency reverses and can be strongly suppressed. This scenario is in agreement with \citet{Verdes-Montenegro2001} who found a deficiency in the gas content of Hickson compact group galaxies. They propose an evolutionary scheme where as a compact group evolves, tidal interactions drive the interstellar medium of galaxies into the intragroup medium of the system. In this line, \citet{Bitsakis2016} suggest that in addition to gas stripping, turbulence and shocks also contribute to suppress star formation in compact groups of galaxies. They correlate the impact of a transient phenomena as shocks, to the frequency of interactions that trigger this mechanism and prevent  interstellar medium of galaxies to relax. 

Given that galaxy groups have a higher probability of repeated interactions than triplets and pairs, it is expected that the star formation suppressing  mechanisms act more efficiently in this systems. This scenario is in agreement with our findings, that show a variation of the specific star formation rate, stellar populations ages and colours of galaxies, which become less star-forming, with older stellar populations and redder colours as the number of members galaxies in the system  increases.

An important factor that may be influencing the properties of galaxies in small systems is the environment at scales larger than those considered for the local isolation of the systems. For this reason, a study of the global environment of small systems of galaxies can add useful insights to the results presented in this paper. This topic will be explored in a forthcoming paper.

The catalogue of small galaxy systems constructed in this work is public available at the Argentine Virtual Observatory home page (http://nova.conicet.gov.ar/).

\section{Acknowledgments}
We thank the referee, Elmo Tempel, for providing us with helpful comments that improved this paper.
This work was supported in part by the Consejo Nacional de Investigaciones Cient\'ificas y T\'ecnicas de la Rep\'ublica Argentina (CONICET) and Secretar\'ia de Ciencia y T\'ecnica de la Universidad Nacional de San Juan. Funding for the Sloan Digital Sky Survey IV has been provided by the Alfred P. Sloan Foundation, the U.S. Department of Energy Office of Science, and the Participating Institutions. SDSS-IV acknowledges support and resources from the Center for High-Performance Computing atthe University of Utah. The SDSS web site is www.sdss.org. SDSS-IV is managed by the Astrophysical Research Consortium for the Participating Institutions of the SDSS Collaboration including the Brazilian Participation Group, the Carnegie Institution for Science, Carnegie Mellon University, the Chilean Participation Group, the French Participation Group, Harvard-Smithsonian Center for Astrophysics, Instituto de Astrof\'isica de Canarias, The Johns Hopkins University, Kavli Institute for the Physics and Mathematics of the Universe (IPMU) / University of Tokyo, Lawrence Berkeley National Laboratory, Leibniz Institut f\"ur Astrophysik Potsdam (AIP),  Max-Planck-Institut f\"ur Astronomie (MPIA Heidelberg), Max-Planck-Institut f\"ur Astrophysik (MPA Garching), Max-Planck-Institut f\"ur Extraterrestrische Physik (MPE), National Astronomical Observatories of China, New Mexico State University, New York University, University of Notre Dame, Observat\'ario Nacional / MCTI, The Ohio State University, Pennsylvania State University, Shanghai Astronomical Observatory, United Kingdom Participation Group, Universidad Nacional Aut\'onoma de M\'exico, University of Arizona, University of Colorado Boulder, University of Oxford, University of Portsmouth, University of Utah, University of Virginia, University of Washington, University of Wisconsin, Vanderbilt University, and Yale University. Mock data used in this work was generated using Swinburne University's Theoretical Astrophysical Observatory (TAO). TAO is part of the Australian All-Sky Virtual Observatory (ASVO) and is freely accessible at https://tao.asvo.org.au. The Millennium Simulation was carried out by the Virgo Supercomputing Consortium at the Computing Centre of the Max Plank Society in Garching. It is publicly available at http://www.mpa-garching.mpg.de/Millennium/.
The Semi-Analytic Galaxy Evolution (SAGE) model is a publicly available codebase that runs on the dark matter halo trees of a cosmological N-body simulation. It is available for download at https://github.com/darrencroton/sage.
\bibliographystyle{mnras.bst}
\bibliography{Bib}{}
 
\label{lastpage}

\end{document}